%% file: main.tex
\documentclass[journal,9pt]{IEEEtran}
\usepackage{amsmath,amsfonts}
\usepackage{algorithmic}
\usepackage{algorithm}
\usepackage{array}
\usepackage[caption=false,font=normalsize,labelfont=sf,textfont=sf]{subfig}
\usepackage{textcomp}
\usepackage{stfloats}
\usepackage{url}
\usepackage{verbatim}
\usepackage{graphicx}
\usepackage[nocompress]{cite}
\usepackage{bm}
\usepackage{xcolor}
\usepackage{hyperref}
\usepackage{booktabs}
\usepackage{ragged2e}
\usepackage{enumitem}

\definecolor{darkred}{rgb}{0.6, 0, 0}
\definecolor{darkgreen}{rgb}{0, 0.5, 0}
\definecolor{darkyellow}{rgb}{0.8, 0.6, 0.1}

\definecolor{michael}{RGB}{18, 179, 149}  
\definecolor{ugcolor}{RGB}{255,0,0}

\begin{document}
\bstctlcite{IEEEexample:BSTcontrol}

\title{Towards Understanding Systems  Trade-offs in Retrieval-Augmented Generation Model Inference}

\author{
    \begin{tabular}{c}
    Michael Shen\textsuperscript{1}, Muhammad Umar\textsuperscript{1}, Kiwan Maeng\textsuperscript{3}, G. Edward Suh\textsuperscript{1,2}, Udit Gupta\textsuperscript{1} \\[0mm]
    \small \textsuperscript{1}Cornell University, 
    \textsuperscript{2}NVIDIA, 
    \textsuperscript{3}Pennsylvania State University  \\[0mm]
    \small \{mts247, mu94, ugupta\}@cornell.edu, esuh@nvidia.com, kvm6242@psu.edu
    \end{tabular}
    \thanks{This material is based upon work supported by the National Science Foundation Graduate Research Fellowship under Grant No. DGE – 2139899.}
\vspace{-0.3in}
}

\newcommand{\note}[1]{\textcolor{blue}{[\small ~#1~]}}


\maketitle

\input{Sections/00-abstract}

\input{Sections/01-introduction}
\input{Sections/02-rag}

\input{Sections/03-methodology}
\input{Sections/04-characterization}

\input{Sections/05-conclusion}

\bibliographystyle{IEEEtran}
\bibliography{references}
\end{document}

%% file: Sections/00-abstract.tex
\begin{abstract}
The rapid increase in the number of parameters in large language models (LLMs) has significantly increased the cost involved in fine-tuning and retraining LLMs, a necessity for keeping models up to date and improving accuracy. Retrieval-Augmented Generation (RAG) offers a promising approach to improving the capabilities and accuracy of LLMs without the necessity of retraining. Although RAG eliminates the need for continuous retraining to update model data, it incurs a trade-off in the form of slower model inference times. Resultingly, the use of RAG in enhancing the accuracy and capabilities of LLMs often involves diverse performance implications and trade-offs based on its design. In an effort to begin tackling and mitigating the performance penalties associated with RAG from a systems perspective, this paper introduces a detailed taxonomy and characterization of the different elements within the RAG ecosystem for LLMs that explore trade-offs within latency, throughput, and memory. Our study reveals underlying inefficiencies in RAG for systems deployment, that can result in TTFT latencies that are twice as long and unoptimized datastores that consume terabytes of storage. 
\end{abstract}

\begin{IEEEkeywords}
Retrieval-Augmented Generation, Information Retrieval, Large Language Models, Natural Language Processing
\end{IEEEkeywords}

%% file: Sections/01-introduction.tex
\section{Introduction}
Since ChatGPT's launch in late 2022, LLMs have found widespread application in research and industry. 
Research in the field has exploded, leading to the development of increasingly more complex models such as GPT-4 and Gemini Ultra. 
The enhanced abilities of language models have enabled researchers to investigate opportunities for personalized learning, advanced analytics, and nuanced human-computer interactions. 

Given their growing adoption, it is crucial to update the knowledge store of LLMs based on user-specific, private data and evolving societal trends. 
To take into account new data, LLMs typically require retraining or fine-tuning. 
However, the growing volume of new information and exponentially increasing LLM sizes~\cite{shoeybi2019megatron} means constant retraining is impractical due to high infrastructure costs~\cite{nlp-training}.

One promising solution to enable LLM ``knowledge'' updates without high training costs is Retrieval-Augmented Generation (RAG) \cite{lewis2021retrievalaugmented,pmlr-v162-borgeaud22a,min2023silo,shi2023replug}. 
RAG enhances LLMs by integrating additional retrieved contexts with a model's generative capabilities. 
Fundamentally, RAG augments models with a database that can be dynamically updated. 
Users' input queries are used to search the database to retrieve relevant contexts; the original input query and retrieved context are then used as input to LLMs to guide the model's response. 
By augmenting input prompts with relevant contexts, RAG has been shown to significantly improve the accuracy and reliability of LLMs~\cite{chen2023benchmarking}. 

While RAG introduces significant modeling advancements to LLMs, it introduces unique challenges to efficient deployment, which warrant further systems characterization and investigation. 
Retrieving input-dependent contexts requires performing similarity searches over large databases that add significant storage and runtime performance overheads.
Consequently, recent efforts have begun looking for ways to optimize RAG performance~\cite{jiang2024piperag,jin2024ragcache,zhang2024accelerating}.
Despite these recent efforts, the field is still in its early stages.
In order to maximize the at-scale performance and efficiency of RAG-based LLMs, we must go beyond end-to-end latency to understand the impact of RAG on Time-To-First Token (TTFT) latency, tail latency, throughput, and memory scalability.

In this paper, we present an initial systems-level characterization of the performance implications of RAG.
To guide our analysis, we construct a taxonomy of RAG systems based on recent RAG-based LLM literature. 
Our taxonomy includes key aspects such as retrieval algorithms (e.g. memory versus performance optimized), mechanisms to integrate retrieval and inference, LLM models (e.g. encoder \& decoder), and runtime parameters (e.g. batching).
Using our taxonomy, we build an extensible framework based on existing open-source implementations of each stage that shows a wide systems' design space based on latency, throughput, storage capacity, and accuracy. 
The main contributions of this work are as follows:

\begin{itemize}
\item We show RAG introduces significant latency overhead to both TTFT latency and end-to-end latency, 
with retrieval making up 41\% of the end-to-end latencies and 45\%-47\% of the TTFT latencies in our setups. 
Furthermore, design decisions on how retrieval and inference stages are integrated (i.e., retrieval stride or frequency of retrieval) expose trade-offs between accuracy and end-to-end inference latency. 
While prior work highlights how frequent retrievals can optimize accuracy, we find that naively re-retrieving can increase the end-to-end latency to nearly 30 seconds, precluding production deployment.

\item We find the choice of a retrieval algorithm has a significant impact on memory requirements and runtime performance, an aspect that must be carefully balanced for efficient RAG system design. 
For instance, memory-efficient retrieval algorithms can require 2.3$\times$ 
less DRAM capacity but achieve a low recall (e.g., 0.65), whereas memory-inefficient retrieval algorithms achieve up to 0.95 recall at the expense of DRAM capacity. 

\item Finally, we show that RAG introduces significant scalability challenges, especially in production-like environments with large datastore sizes and high query volumes.
As the datastore size grows from 1 million to 100 million chunks, the throughput of the retrieval stage degrades by up to 20$\times$.
Furthermore, billion-scale datastores can require TB-scale memory capacity; while memory-efficient retrieval algorithms can reduce the memory capacity overhead, they scale poorly compared to memory-inefficient retrieval algorithms at comparable recall rates (yielding up to 2.2$\times$ lower throughput at larger batch sizes).
\end{itemize}

%% file: Sections/02-rag.tex
\IEEEpubidadjcol
\section{Retrieval-Augmented Generation}
Traditional pretrained (decoder-only) language models for text generation work by processing an input query and then generating output tokens autoregressively, leveraging the attention mechanism that allows the model to accurately weigh the importance of different parts of the input and previously generated output tokens.  

\begin{figure}[t]
    \centering
    \includegraphics[width=\linewidth]{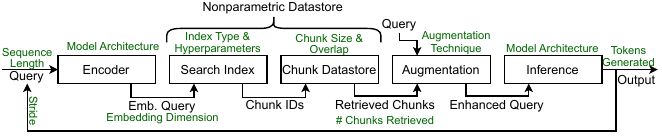}
    \vspace{-2em}
    \caption{Taxonomy of the online RAG inference workflow and various adjustable hyperparameters (highlighted in green), that can be used to modify and enhance the workflow.}
    \label{fig:fig1}
    \vspace{-1em}
\end{figure}

RAG differs from this traditional workflow by incorporating a dynamic, non-parametric datastore that is queried for relevant context before the inference model processes the query (as seen in ~\autoref{fig:fig1}). 
This workflow can be divided into two distinct stages: 
\begin{itemize}[leftmargin=*]
    \item \textbf{Offline Datastore Preprocessing}: where the datastore along with efficient indices for chunk retrieval are constructed.
    \item \textbf{Online Inference}: where the constructed datastore indices are used to enhance the traditional LLM workflow. 
\end{itemize}
In this section, we describe the essential design components of both of these stages and analyze how they can affect the performance of RAG models. 

\subsection{Offline Datastore \& Index Construction}
\IEEEpubidadjcol
The primary objective of the offline stage is to prepare the non-parametric datastore for efficient information retrieval during the online inference stage. 
This involves taking the original documents from the datastore, splitting them into chunks (of a certain size, with or without overlap), and assigning them IDs.
Relevant chunks will need to be retrieved for online queries, which can be done either via dense (semantic) retrieval in vector space, or sparse retrieval using words/tokens. 
For the purposes of this study, we focus on dense retrieval. 
During the offline phase, each chunk is embedded into a vector using an encoder model.
These vectors are used during the online phase (Section~\ref{sec:bg_online}) to retrieve chunks that are closest to the user query using a nearest neighbor search.
For an efficient search, retrieval indices are built over the vectors for a fast Approximate Nearest Neighbor (ANN) search, followed by a chunk ID lookup to retrieve the chunks. 

The design and implementation of this non-parametric datastore include important design decisions that can impact the behavior of the retrieval process and the end-to-end online RAG inference. 
Salient design decisions include:

\begin{itemize}[leftmargin=*]
\item \textbf{Size}: The amount of content within the datastore and the resulting memory footprint impact the potential search time required for the RAG model to find relevant chunks.

\item \textbf{Content}: Datastore contents can comprise of chunks, chunk vectors, and optionally chunk metadata for additional filtering. 

\item \textbf{Index Type}: Advanced indexing strategies we consider for this characterization study, such as 
Hierarchical Navigable Small World (HNSW)~\cite{malkov2020efficient} and Inverted File (IVF)~\cite{zobel2006inverted}, introduce trade-offs that impact search efficiency, accuracy, and memory storage. 
HNSW is a state-of-the-art algorithm for efficient ANN searches that utilizes a graph structure for rapid searches, with similar vectors as proximal nodes.
It uses a multi-layered graph where higher layers have fewer nodes to efficiently guide searches to relevant areas. 
Lower layers, containing more nodes, allow for detailed search within those areas.
IVF clusters similar data together, enabling focused searches within these clusters rather than the entire dataset. 
Additionally, relevant search-time parameters such as \texttt{efSearch}, which dictates how extensively nearest neighbors are explored at query time in HNSW, and \texttt{nProbe}, which controls how many unique clusters are searched at query time for IVF, can be adjusted to balance throughput and accuracy trade-offs.

\item \textbf{Storage Location}: Datastores with large memory footprints require alternatives to in-memory storage, such as on disk or in remote servers, that can lead to additional communication overhead.
\end{itemize}

\subsection{Online Inference}
\label{sec:bg_online}
When a query is first passed into a RAG model it is encoded. The query is then used to index and search into the non parametric retrieval indices for the IDs associated with the $k$ approximately nearest neighbors. With the chunk IDs retrieved, a lookup is done to retrieve the text chunks for the corresponding IDs in the datastore.   

\color{black}
Once chunks from the non-parametric datastore are retrieved, they can be re-ranked to select the most relevant chunks or to assign weights to chunks for inference \cite{shi2023replug}. 
The re-ranking can be simple (e.g. using similarity scores), or advanced (e.g. using neural networks to better assess chunk relevance~\cite{sbert}).

With the chunks retrieved, they are integrated into the inference process through a variety of techniques such as prepending to the input query\cite{shi2023replug} or cross-attending to chunk embeddings during inference\cite{pmlr-v162-borgeaud22a}.
Combining multiple chunks can happen via simply concatenating them together, or other techniques (e.g. combining output inference probabilities after separately processing each chunk\cite{shi2023replug}).
Similar to the encoder model parameters, inference model parameters such as size, architecture, and context/output length can greatly impact the RAG workflow and performance.
In state-of-the-art RAG systems~\cite{riclm}, retrieval striding is applied, where multiple retrieval iterations occur per inference pass, fetching new documents every $s$ generated tokens to enhance output accuracy. \color{black}

%% file: Sections/03-methodology.tex
\section{Experimental Setup}
Based on our taxonomy, we design one possible RAG-based LLM pipeline to evaluate performance trade-offs, utilizing representative state of the art open-source models and retrieval indices. \color{black}

\textbf{Datasets and models.} We use a BGE Large encoder model to encode queries at run-time~\cite{bge_embedding}.  
For inference, we use the open-source 9B GEMMA 2 \cite{gemma_2024} model.\color{black}

Our baseline configuration assumes an input sequence length of 512 tokens and a generated output of 128 tokens, based on the average length seen in production systems~\cite{patel2023splitwise}. We have a single retrieval per query (i.e., no retrieval stride).
Following recent work optimizing cloud efficiency for LLM inference by balancing latency and throughput, we study a scenario where each stage runs at a distinct batch size in a disaggregated way to balance systems' efficiency.
Given our models and hardware systems for encoding, retrieval, and inference, we set a default batch-size of 32 for all stages.\color{black}

\textbf{Retrieval indices.} 
We construct our retrieval indices using a 100M document chunk subset of Common Crawl \cite{piktus2021web}.
Each document chunk is made up of 100 tokens, encoded into a vector. \color{black}
We explore three different dense retriever indices: 
an accuracy-oriented one (HNSW-SQ: HNSW with 8-bit scalar quantization, 128 links per graph node, \texttt{efConstruction}=200, \texttt{efSearch}=128), a storage-oriented one (IVF-PQ: IVF with 256-dim 8-bit product quantization 16384 cells, \texttt{nProbe}=64), and a balanced accuracy/storage one (IVF-SQ: IVF with 8-bit scalar quantization 16384 cells, \texttt{nProbe}=64).
For retriever performance and recall analysis, we leverage queries from the TriviaQA-test dataset \cite{joshi2017triviaqa}.

During retrieval, we retrieve 20 approximately nearest chunks.
After each retrieval, we prepend the 2 nearest chunks from the 20 (obtained via re-ranking using inner-product distance with the query vector) to the input query, to form the prompt for LLM generation.

\textbf{Hardware.} For retriever experiments, we use 32 cores of an Intel Xeon Silver 4316, and the FAISS \cite{douze2024faiss} ANN library.
For Transformers, we leverage an NVIDIA A6000 ADA GPU, using the HuggingFace \cite{hftransformers} library with vLLM \cite{kwon2023efficient}, under FP16 precision.\color{black}


%% file: Sections/04-characterization.tex
\section{Experimental Results}

\begin{figure*}[ht]
    \centering
    \includegraphics[width=\linewidth]{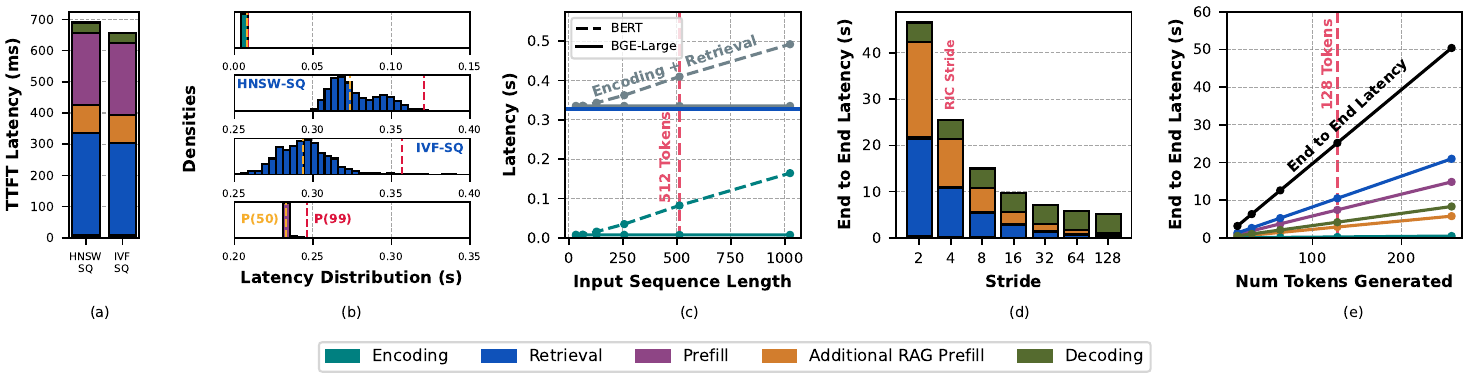}
    \vspace{-1.5em}
    \caption{
    (a) Breakdown of TTFT latency for RAG using two retrieval algorithms (HNSW-SQ and IVF-SQ) with comparable accuracy, highlighting that RAG significantly increases latency on top of prefill and decoding.
    (b) Analysis of RAG component tail latencies, demonstrating that unique stages of RAG, such as retrieval, experience much longer tails.
    (c) Examination of the continued impact of retrieval in comparison to varied encoding sequence lengths. 
    (d) Evaluation of various stride lengths and their impacts, illustrating how aggressive strides can substantially increase the end-to-end RAG latency. We highlight a stride of 4 for optimizing perplexity seen within prior works~\cite{riclm}    
    (e) Examination of the impact of varied numbers of generated tokens on all RAG components with a stride length of 4, showing a substantial increase in latency.
    \color{black}
    } 
    \label{fig:fig2}
    \vspace{-1.5em}
\end{figure*}
\color{black}
In this section, we explore the bottlenecks and trade-offs associated with introducing retrieval augmentation into a traditional language model pipeline.
We also study trade-offs within the information retrieval stage. 
Note that our study is not exhaustive. More comprehensive RAG systems (as hinted in our taxonomy) entail a larger range of design choices that shall be studied in future work.
We enlist and describe our findings as the following takeaways.

\textbf{Takeaway 1: Compared to baseline LLM inference, RAG-based LLMs introduce additional overheads that prolong TTFT inference latency.} \color{black}
Figure~\ref{fig:fig2} shows the TTFT latency of RAG-based LLMs using two different retrieval indices that achieve comparable retrieval accuracy: HNSW and IVF with scalar quantization (SQ). 
While prefill and decoding stages are standard in LLM inference, RAG introduces three additional stages: encoding, retrieval, and additional prefill steps for retrieved contexts.
Figure~\ref{fig:fig2}(a) shows that these RAG stages nearly double the TTFT latency, from 495ms to 965ms, compared to the baseline LLM.
In both scenarios, using either HNSW-SQ or IVF-SQ, the most significant portion of the overhead comes from the retrieval stage which accounts for nearly 35\% of the total TTFT latency.
Figure~\ref{fig:fig2}(c) shows that even as the input sequence increases to 1000 tokens, retrieval continues to be dominant compared to the encoding stage's latency.

In addition to average latency, Figure~\ref{fig:fig2}(b) illustrates the impact of RAG on the tail latency of each stage. 
While the gap between p99 and p50 latencies for encoding and prefill stages is minimal, we observe orders of magnitude higher tail latencies for retrieval.
With no co-located workloads and interference, the gap between p99 and p50 latency for HSNW-SQ and IVF-SQ is 50$ms$ and 60$ms$, respectively.
Given the TTFT for production workloads is on the order of a few hundred milliseconds~\cite{patel2023splitwise}, an additional tail latency of about 50$ms$ represents a significant overhead.
Minimizing TTFT is crucial for maintaining an acceptable user experience.
The significant overhead introduced by retrieval stages affects both average TTFT and tail latencies, resulting in performance variability that impacts achievable Quality-of-Service. 
This, in turn, poses challenges for effective resource allocation and workload scheduling strategies in production environments.

\textbf{Takeaway 2: In addition to TTFT, design decisions on how RAG is integrated into LLM inference pipelines (i.e., retrieval stride or re-retrieval frequency) have nearly two orders of magnitude impact on end-to-end latency}. 
Figure~\ref{fig:fig2}(d)-(e) illustrates the overall inference latency when integrating retrieval into end-to-end RAG-based LLM pipelines. 
Adjusting the re-retrieval stride determines the frequency at which new context is retrieved to generate output sequences.
With an extremely aggressive stride length of 4, the end-to-end latency increases to nearly 30 seconds. 
In this case, RAG unique components account for nearly 97\% of the end-to-end latency; retrieval and additional prefill overheads (due to added contexts and additional retrievals) account for 36\% and 45\% of the latency respectively.
Figure~\ref{fig:fig2}(e) shows that this latency overhead also grows with the number of output tokens generated, due to the additional prefill stages from re-retrieval.
In production scenarios, frequently retrieving context is prohibitively expensive.
As such, the benefits of retrieval augmentation come at a significant performance cost, and RAG pipelines need to be optimized to balance accuracy \textit{and} performance to become truly practical.

\vspace{-1ex}
\begin{table}[H]
    \centering
    \caption{Comparison of various retrieval index metrics.}
    \vspace{-1ex}
    \label{tab:tab1}
    \begin{tabular}{lrrr}
        \toprule
         & HNSW-SQ & IVF-PQ & IVF-SQ \\
         \midrule
         Latency (s) & 0.31--0.81 & 0.29--2.19 & 0.17--1.72 \\
         Throughput (QPS) & 25--319 & 27--116 & 47--148 \\
         Recall & 0.87 & 0.61 & 0.86 \\
         Storage (GB) & 166 & 23 & 71 \\
        \bottomrule \\
    \end{tabular} 
    \scriptsize    \parbox{.86\linewidth}{\emph{This illustrates the complexity of the design space and the challenge in selecting an optimal index. Latency and throughput are assessed with batch sizes ranging from 8 to 256, highlighting the higher throughput efficiency of IVF indices at smaller batch sizes and HNSW indices at larger batch sizes.}}    \normalsize
\end{table}
\vspace{-1ex}

\begin{figure*}[ht]
    \centering
        \includegraphics[width=\linewidth]{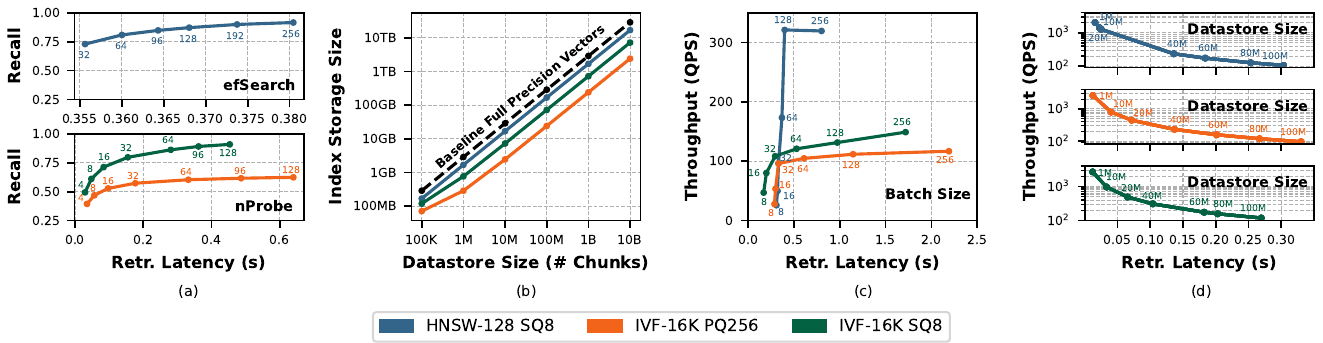}
    \vspace{-1.5em}
    \caption{
    (a) Analysis of the trade-off between recall and latency by adjusting index-specific search-time hyperparameters, showing that higher recall often comes at the cost of lower latencies 
    (b) Illustration of the significant increase in index storage size for datastores as the number of chunks grows, highlighting the substantial memory footprint for larger datastores.
    (c) Comparison of throughput versus latency for various retrieval batch sizes from 8 to 256 for, showing how HNSW indices can scale better comparatively to IVF indices. 
    (d) Breakdown of retrieval throughput vs. latency for indices of varying datastore sizes, indicating the inability of datastores to effectively scale as larger datastores are constructed.
    }
    \label{fig:fig3}
    \vspace{-1em}
\end{figure*}

\textbf{Takeaway 3: Scaling the datastore size may necessitate a change from using a memory-optimized retrieval algorithm instead of an accuracy-oriented algorithm, introducing a non-continuous trade-off between accuracy and latency.}
Figure~\ref{fig:fig3} evaluates the impact of different retrieval algorithms and configurations in terms of recall, latency, throughput, and index memory usage. 
Accuracy-oriented retrieval indices (i.e., HNSW-SQ) achieve high recall at the expense of larger index sizes; for instance, a datastore size of 1 billion chunks incurs 1TB of memory capacity overhead (Figure~\ref{fig:fig3}(b)).
While HNSW-SQ has a logarithmic search complexity and good recall due to 8-bit vector elements, its memory requirement is still large because bidirectional links need to be stored between graph nodes.
Given these large datastore sizes, retrieval search is typically run on general-purpose CPUs as opposed to memory-limited AI co-processors like GPUs.

Compared to the HNSW retrieval algorithm, IVF-based retrieval algorithms are more memory efficient. 
For instance, IVF-PQ and IVF-SQ incur 7.2$\times$ and 2.3$\times$ less memory usage than HNSW-SQ, respectively.
However, while IVF-SQ can achieve comparable recall to HNSW-SQ, IVF-PQ only achieves the maximum recall of $\approx$0.6.
In addition to lower memory usage, IVF also allows trade-offs between retrieval latency and recall; IVF indices can achieve retrieval latencies of as low as 30$ms$ (i.e., 10$\times$ lower than HNSW-SQ) at the expense of low recall (Figure~\ref{fig:fig3}(a)).

\textbf{Takeaway 4: Despite the memory challenges inherent at large scales, RAG models may also struggle to maintain high throughput if not designed effectively for production environments, where exceptionally large datastores and high query volumes are anticipated.} \color{black}
While prior works optimizing the performance of retrieval algorithms~\cite{douze2024faiss} and RAG systems~\cite{zhu2024accelerating} focus primarily on latency, in production environments, throughput is crucial to consider for at-scale deployment.
Figure~\ref{fig:fig3}(c) shows how in addition to memory efficiency trade-offs, the throughput of retrieval algorithms varies significantly as we vary the batch size.
For instance, while HNSW-SQ achieves over 300 QPS with larger batches above 128, IVF-SQ and IVF-PQ can only achieve a throughput of 150 and 110 QPS, respectively.
Serving multiple requests in batches is important for inference to achieve high resource utilization.
The FAISS library schedules one thread per search query and as we schedule larger batch sizes with fixed hardware resources, searches are serialized.
As a result, IVF indices scale worse than HNSW-SQ; IVF indices require much more computation for each search query, whereas HNSW indices can take advantage of work stealing much better.

In addition to achieving high throughput with batching, retrieval in production environments must consider the impact of scaling datastore sizes on performance.
Figure~\ref{fig:fig3}(d) illustrates that as the datastore size grows from 1 million to 100 million chunks, throughput decreases by up to 20$\times$ for our tested search indices. 
100M datastores are considered relatively small. 
When datastores scale to the billions scale, as seen in~\cite{pmlr-v162-borgeaud22a}, the latency and throughput penalty will be significant. 
This observation illustrates the need for intelligent retrieval algorithm choices, and improved retriever designs (e.g. distributed retrieval over multiple nodes with split datastores). This also highlights how challenges in implementing at-scale indices are multifaceted rather than straightforward, with multiple systems considerations arising in the design and scaling of search indices. \color{black}

%% file: Sections/05-conclusion.tex
\section{Conclusion}
In this paper, we present a taxonomy of RAG-based LLMs and conduct a series of case studies to investigate the systems characteristics of the RAG pipeline. Our experimental results reveal a diverse design space and highlight opportunities for system optimizations for efficiently deploying RAG models, particularly in the context of information retrieval. Future work will involve a more thorough exploration of the design space (e.g., sparse retrievers, query transformation techniques, re-ranking algorithms) and an end-to-end evaluation of the entire RAG pipeline. Additionally, this work underscores the need for designing systems that more optimally support RAG inference for efficient and accurate deployment.